\documentclass{IEEEcsmag}

\usepackage[colorlinks,urlcolor=blue,linkcolor=blue,citecolor=blue]{hyperref}

\jvol{XX}
\jnum{XX}
\paper{8}
\jmonth{May/June}
\jname{IEEE Computer}
\pubyear{2022}

\begin{document}

\title{Guidelines for Cultivating a Sense of Belonging to Reduce Developer Burnout}

\author{Bianca Trinkenreich}
\affil{Colorado State University and Oregon State University}

\author{Marco Aurelio Gerosa}
\affil{Northern Arizona University}

\author{Anita Sarma}
\affil{Oregon State University}

\author{Igor Steinmacher}
\affil{Northern Arizona University}

\maketitle

\chapterinitial{Burnout affects software developers' mental and physical well-being and contributes to their turnover, which has generated strong industry concerns. Our previous research showed that lack of belonging is associated with higher levels of burnout in software developers. A sense of belonging is also key to making people more resilient and have better job satisfaction and well-being.
In this paper, we revisit our recent research on these topics to offer guidelines on promoting a sense of belonging in software development teams, ultimately improving developers' well-being.}

\section{Resilience through Belongingness}

Resilience refers to the ability of an individual to adapt and recover from difficulties, challenges, or adverse situations. It is linked to enhanced mental health and well-being~\cite{fletcher2013psychological}. Establishing a sense of belonging within a team, group, or network of supportive individuals can lay a strong foundation for resilience development~\cite{blanchard2007developing}. When individuals feel a strong sense of belonging, they are more likely to experience increased job satisfaction, engagement, and overall well-being. Prior research suggests that high belongingness within a team can contribute to higher levels of job satisfaction and lower rates of burnout \cite{trinkenreich2023model}.

Burnout threatens well-being and encompasses feelings of exhaustion on physical, emotional, and cognitive fronts. This has likely influenced the decision to exit organizations in all industries. Understanding burnout is especially important to software developers, who are more likely to feel fatigued, anxious, experience burnout, and stressed than those who perform mechanical tasks~\cite{nayak2014anxiety}. 

Sense of belonging has been shown to prevent or reduce burnout by increasing job satisfaction \cite{trinkenreich2023model}. By increasing the sense of belonging, one may feel less isolated and find a toolset to increase well-being. In this paper, we bring a sense of belonging perspective on burnout in software development. 

\section{Fostering a sense of belonging}

Belongingness measures to what extent people feel like they are part of a group or fit into a certain group~\cite{hagerty1995developing}. According to Maslow~\cite{maslow1943theory}, it is a basic human need. Hagerty and his team~\cite{hagerty1995developing} go further, saying that belongingness is a unique mental health idea, different from things like loneliness or getting support from others. They define two dimensions for belonging: (i) feeling valued and accepted and (ii) fitting in as your qualities match with the setup or place. Belongingness leads to a positive, supportive work atmosphere. 

While the sense of belonging is important for everyone, it is especially relevant for underrepresented groups, and addressing these needs is key to promoting personal well-being and effective functioning. Research on belongingness in STEM education has highlighted challenges faced by minority groups, such as women in mathematics~\cite{good2012women} and Latinx individuals dealing with racially hostile environments, underscoring the struggles these groups encounter in terms of belonging.

Nonetheless, the concept of belongingness holds a comparable significance for software developers. Software development occurs in different ecosystems with different incentives and work structures. Two main ecosystems include corporate settings and Open Source Software (OSS).

In corporate settings, software developers are remunerated for their efforts and backed by a centralized Human Resources team. Conversely, OSS projects depend on input from a diverse community that mainly interacts online and encompasses individuals with varied motivations and compensation structures. This community involves volunteers and contributors who may be retained and funded by separate enterprises and foundations.
 
Hence, we investigated belongingness in both the IT industry and OSS.

While belongingness can motivate software developers and lead to disengagement when lacking, we currently do not have clear guidance to foster a sense of belonging within a team. In the following section, we discuss actions organizations can take to cultivate belongingness in a software team.

\section{Deciphering Belongingness for Software Developers}

\begin{figure*}[htb]
\centering
\includegraphics[width=0.95\textwidth]{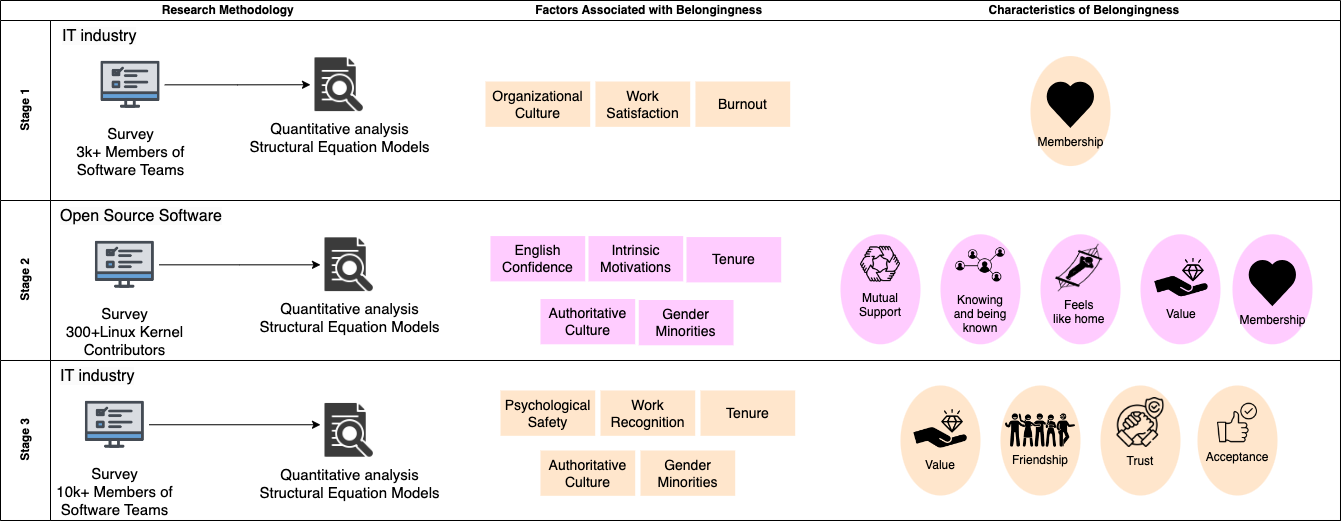}
\caption{The methodology employed to study characteristics and factors associated with belongingness in software development teams. The result from Study 1 was published at ICSE-SEIP 2023 \cite{trinkenreich2023model}, the results from Study 2 were published at ICSE 2023 \cite{trinkenreich2023belong} and the results of Study 3 were published at ICSE 2024 \cite{trinkenreich2024belong}. 
} 
\label{fig:methodology}
\end{figure*}

To establish guidelines for enhancing belongingness among software developers, we conducted three studies
within different software development environments, including open-source and closed-source environments, illustrated in Figure \ref{fig:methodology} and summarized in Table \ref{tab:studies}.

\begin{table*}[htb]
\centering
\caption{Summary of studies included in this paper}
\label{tab:studies}

\begin{tabular}{p{5cm}
                p{5cm}
                p{5cm}}
\hline
{Study 1: Case Study} & {Study 2: Case Study} & {Study 3: Case Study} \\ 
\hline 
{May 2023} & {May 2023} & {Oct 2023} \\ 
\hline
{How does organizational culture impact inclusiveness, belongingness, team climate, work satisfaction, and developer burnout?} & {How does a sense of virtual community develop in Open Source Software projects?} & {How are work appreciation and psychological safety related to the sense of belonging in software development teams?} \\
\hline 
{Approach: Analysis of 3,281 survey responses from software delivery team members across diverse projects at a proprietary software company} & {Approach: Analysis of 318 survey responses from Linux Kernel contributors} & {Approach: Analysis of 10,781 survey responses from software development team members across diverse projects at another proprietary software company} \\
\hline 
{Key Finding: Organizational Culture, Climate for Learning, Sense of Belonging, and Inclusiveness are positively associated with Work Satisfaction, which in turn is associated with Reduced Burnout.} & Key Finding: {Intrinsic motivations positively relate to the sense of virtual community, while gender minorities perceive a weaker virtual community compared to men. Power distance negatively affects the sense of virtual community, while English proficiency and tenure have a positive association with it.} & Key Finding: {Psychological safety and work appreciation positively correlate with a sense of belonging to the team. Gender differences are observed, with women feeling less belonging. Power distance has a negative association with belonging, whereas tenure shows a positive association.}\\

\hline
\end{tabular}

\end{table*}

To explore the connection between belongingness and burnout, we conducted Study 1 by analyzing software development teams at a global tech company. We found a positive association between organizational culture and belongingness. More interestingly, \textit{belongingness was found to be positively linked to job satisfaction and inversely related to burnout.}

Motivated to discover additional factors that impact the sense of belonging, to delve into further aspects of belonging beyond membership, and to reveal the dynamics of belonging in different ecosystems, we conducted two other studies. Study 2 \cite{trinkenreich2023belong} was conducted within a large Open Source Software project, while Study 3 \cite{trinkenreich2024belong} focused on software development teams working on another big company that develops proprietary software solutions for clients. 
Studies 1 and 3 happened in large organizations where the workforce is financially compensated and produces proprietary software. In contrast, Study 2 was conducted in an Open Source ecosystem comprising a mix of volunteers and paid contributors that are financially compensated.

Study 2 investigated belongingness within an OSS team \cite{trinkenreich2023belong}, utilizing attributes from the Sense of Virtual Community \cite{blanchard2007developing} framework, including (membership, mutual support, knowing and being known, feeling like home, and feeling valued). We explored associations between belongingness and intrinsic motivations, English proficiency, tenure in the team, authoritative cultural traits (power distance), and gender.
This study was based on the results from the survey with 318 contributors from the Linux Kernel. It showed that the sense of virtual community tends to be more pronounced among individuals motivated by kinship and who have been contributing to the project for a long time. It is characterized by a sense of membership and a strong emotional bond with the group \cite{blanchard2007developing}. This feeling is further manifested in the perception that members hold importance for both the collective and each other, which is facilitated by mutual support, sharing of personal identity, and getting to know the identities of fellow members.

We then embarked on Study 3 within the software development teams that produce proprietary software at another global tech company  \cite{trinkenreich2024belong}. Here, teams interact both online and in person, and we employed the characteristics of belongingness derived from the Sense of Belonging instrument (SOBI) \cite{hagerty1995developing}, encompassing friendship, trust, acceptance, and value recognition. We brought the Self-Determination Theory~\cite{ryan2017self}, which suggests that psychological safety, recognition, and belonging are interconnected psychological needs that promote motivation and engagement, leading to a positive and supportive work environment. In this study, we investigated the correlation between belongingness and psychological safety, work appreciation, tenure at the company, authoritative culture (country culture dimension of power distance), and gender. Results showed that women and those living in countries with high power distance show a relatively low sense of belonging. On the opposite side, those who have more experience feel more belonging than novice members. Nevertheless, when examining the different roles (leaders and non-leaders), gender has a different impact on feelings of belonging. Women who do not occupy leadership positions feel less belonging than those who are leaders \cite{trinkenreich2024belong}.

\begin{figure*}[htb]
\centering
\includegraphics[width=0.80\textwidth]{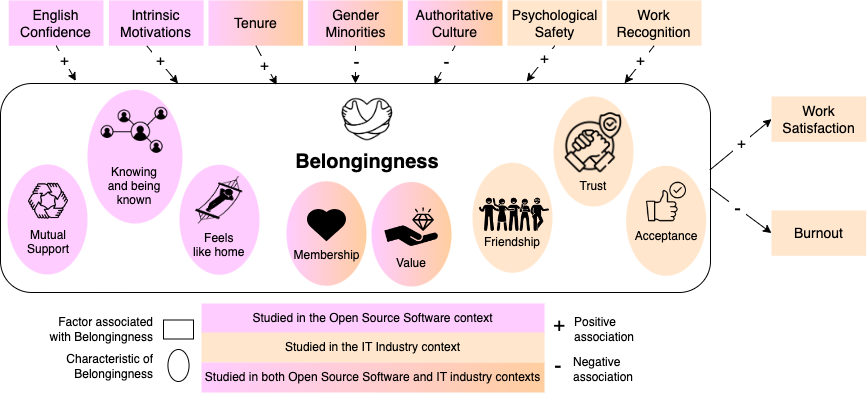}
\caption{Characteristics and Factors Associated with Belongingness. 
Squares depict factors correlated with belongingness, while ovals symbolize the attributes of belongingness probed in the various studies. Arrows signify statistically significant associations uncovered throughout the studies. A positive association signifies an increase, whereas a negative association indicates a decrease.}
\label{fig:belongingness}
\end{figure*}

In the present paper, we organized the characteristics of belongingness and the factors that we investigated as associated with belongingness, as shown in Figure \ref{fig:belongingness}.

\begin{itemize}
\item Belongingness characteristics studied in the IT Industry: being able to \textsc{trust} other people from the team, feeling \textsc{accepted} and \textsc{valued} by co-workers, having \textsc{friendship} and a good relationship with other members of the team
\item Belongingness characteristics studied in Open Source Software: feeling as a \textsc{member} of the team, being \textsc{known by others} and \textsc{knowing who to ask} for help in case of need, feeling \textsc{valued} and perceiving the team is \textsc{like home}.
\item Factors that are associated with belongingness, studied in the IT industry: work recognition (including compensation, leadership recognition, and perceived to make a difference to the company); and psychological safety (including being able to be yourself, feeling safe to share bad news with the team, reporting improper behaviors such as discrimination or harassment, and speaking up and taking risks)
\item Factors associated with belongingness, studied in Open Source Software: intrinsic motivations (including kinship and fun) and English confidence (including both spoken and written in technical and non-technical communication).
\item Demographic factors associated with belongingness, studied both in the IT industry and in Open Source Software: living in an authoritative culture (high score in Power Distance country culture \cite{hofstede2001culture}), gender and tenure.
\end{itemize}

\section{Improving Belongingness}

\begin{figure*}[htb]
\centering
\includegraphics[width=1\textwidth]{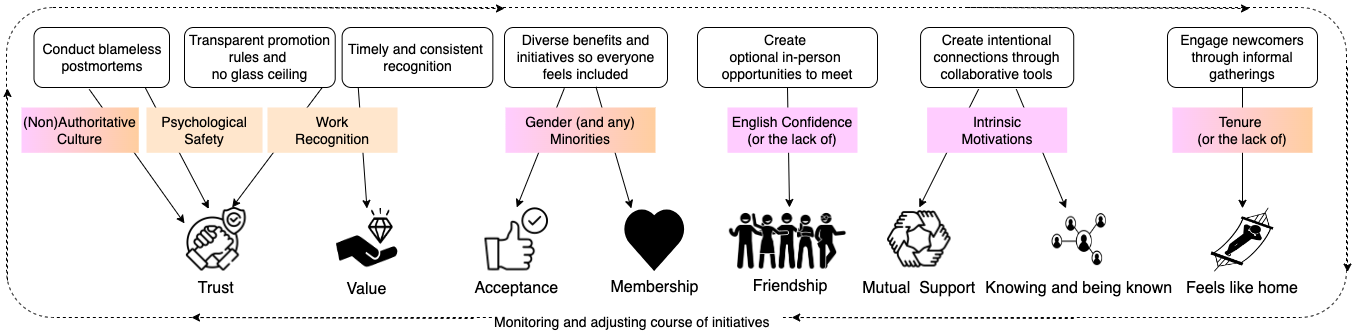}
\caption{Strategies to Cultivate Belongingness} 
\label{fig:strategies}
\end{figure*}

Based on the evidence found during quantitative analysis of belongingness in Open Source Software and IT industry contexts, we propose guidelines for leaders to cultivate belongingness in the team and thus, reduce burnout. All guidelines presented here are backed by evidence previously collected or supported by the literature. In Fig.~\ref{fig:strategies}, we present how the different guidelines' strategies can support achieving the different dimensions of belongingness through the factors (squares) we found associated with the sense of belonging.

\textbf{Timely and consistent recognition.} Positive leadership feedback is crucial for acknowledging employees, fostering belongingness, and reducing burnout \cite{trinkenreich2023model}. Strive for a workplace culture in which individuality is both noticed and \textsc{valued}\cite{gartner_belonging_2022}. Crafting meaningful feedback can be complex, requiring consistency, timeliness, tailoring, and appropriateness. Consistency involves regular recognition, ideally quarterly, to maintain impact without overwhelming. Timeliness ensures prompt acknowledgment to sustain the significance of accomplishments. Tailoring recognition of an employee's actions is important, as generic thanks can be discouraging. Leaders can articulate the impact of individual efforts on the software quality or the team's sustainability. For instance, leaders can highlight project achievements, such as delivering software on time or delivering code that meets high-quality standards by being bug-free. Furthermore, recognition should go beyond technical accomplishments, encompassing actions that support team dynamics. ---e.g., acknowledging instances where a senior developer paired with a junior developer helped the junior's growth. Appropriateness involves using suitable approaches and communication channels. People prefer various forms of recognition---public praise, private messages, or tangible rewards. Understanding preferences and using multiple channels, like email, phone, or in-person interactions, helps foster a sense of \textsc{value} and connection. Tangible expressions of appreciation include job security, transparent career advancement opportunities, and social welfare benefits for workers.

\textbf{Transparent promotion rules and no glass ceiling:} Establishing a workplace culture centered on office politics and internal rivalries can erode trust and leave employees feeling that their workplace lacks fundamental fairness. Healthy competition can be beneficial, but overly competitive atmospheres can erode camaraderie and a sense of belonging. Even leaders committed to creating an equitable workplace must continually challenge themselves to identify areas where they may fall short. Weak leadership and favoritism can tarnish efforts to build an inclusive workplace, whereas exceptional leaders will cultivate an inclusive environment for everyone. The promotion rules need to exist and be followed. Promotions are the organization’s opportunity to demonstrate its \textsc{values} and a key area where your organization can fail to live up to its stated ideals. For example, if leaders tout the importance of women workers but keep overlooking them for promotions when their skills match technical team leaders' or management positions, \textsc{trust} is lost. Those women will feel their coding skills are undervalued and they do not belong in your organization. Hence, opening the criteria to advance to higher-level company positions or becoming an OSS maintainer is crucial \cite{trinkenreich2021pot}.

\textbf{Diverse benefits and initiatives so everyone feels included:}
Providing employees or OSS contributors with benefits and initiatives that honor their unique contributions to the organization demonstrates that business success is directly linked to whether or not employees feel like they are \textsc{accepted} and belong. Gartner's research \cite{gartner_belonging_2022} indicates that offering tangible benefits that cater to diverse needs, such as flexible work schedules and emotional well-being programs, enhances feelings of inclusion by as much as 38\%, especially for minorities. When considering diverse needs to create strategies of benefits and initiatives, the organization is bringing a message that there is not ``one size fits some" with the expectation that everybody else squeezes in. These displays of appreciation convey your genuine concern for their individual requirements both within and outside the workplace. Holding celebratory events about underrepresented groups (e.g., Black Women’s History Month, Chinese New Year's Eve) is a highly effective strategy to cultivate belonging for a diverse set of people \cite{gartner_belonging_2022}.

\textbf{Create intentional connections through collaborative tools} as online interest groups for members, chat rooms, instant messaging, and discussion forums can be provided to encourage team involvement that happens in virtual settings. Members of the team can use those shared spaces to ask questions and obtain \textsc{mutual support}, but also to build social capital and develop relationships, find people with similar schedules and interests for pair programming \cite{sedano2016sustainable}, being able to \textsc{know other members and being known by others too.} Peer programming and shared spaces are also important for contributors intrinsically motivated by kinship to work ``together" on issues (e.g., deploying code together), discussing and collaborating on similar interests. Leaders should manage pull requests, mailing lists, or group messages and guarantee that members' posts are not missed and that the communication adheres to the code of conduct or company policies.

\textbf{Conduct blameless postmortems} to improve \textsc{psychological safety} and build trust across the team. Altering behavior and work patterns shift culture. Teams can establish practices for a non-authoritative and generative culture that encourages information sharing and trust, increases the sense of belonging and reduces burnout \cite{trinkenreich2023model}. This involves applying an organizational culture model focused on generative behaviors that can be done during Agile retrospectives, training messengers to bring bad news and training the group to learn from failure, fostering an environment where risk-taking is accepted, where the messenger is not punished. Members address issues early and emphasize inquiry over blame.

\textbf{Create optional in-person opportunities to meet} as meetups, group meetings, and conferences. 
In-person interactions can significantly boost \textsc{friendship} through social interactions. 
DevOps Days\footnote{\url{https://devopsdays.org/}} and The Linux Foundation gatherings\footnote{\url{https://events.linuxfoundation.org/}} are examples of global events where developers share ideas, learn, and collaborate. While face-to-face interactions in these events are easier for those with English proficiency, others can also actively engage in conversations in a friendly, safe environment to develop their language skills and gain the confidence to communicate effectively in English.

Additionally, spontaneous office exchanges and feedback loops amplify the feeling of belonging \cite{tkalich2022happens}. Following the pandemic, several software developers shifted to remote work, engaging primarily through virtual tools, entirely or partially. Encountering one another, even if not on a daily basis and particularly without the compulsion of an on-site work arrangement, contributes to cultivating a sense of belonging and alleviating job burnout.

\textbf{Engage newcomers through informal gatherings}
 Utilize ice-breaker questions, focusing on safe and casual topics like favorite food or hobbies and inviting opinions and perspectives into the conversation, especially during the first meetings of a software development project. This approach cultivates a relaxed atmosphere that helps the newcomer feel at ease within the group, like \textsc{feeling at home}. By doing so, both the newcomer and the existing members are prompted to engage, familiarize themselves with others, and share their own backgrounds and preferences. This interaction paves the way for exchanging cultural, geographic, and behavioral insights, ultimately promoting diverse viewpoints and strengthening the bonds among team associates.

\textbf{Monitoring and adjusting course of initiatives.} While numerous variables can influence these initiatives' outcomes, evaluating their effectiveness consistently remains essential. This can be achieved through regular assessments, such as conducting brief monthly surveys with three or four specific questions designed to gauge various aspects of a sense of belonging and employee burnout. These assessments serve a vital purpose in several ways: (a) Continuous Monitoring: Ensure that you have an ongoing pulse on the well-being of your employees and have a proactive measure to detect issues early, even before they escalate; (b) Quick Response: Swift responses to emerging concerns when trends or issues are identified to prevent potential negative consequences; (c) Data-Driven Decisions: Enable leaders to tailor their interventions and initiatives based on the feedback received from their workforce; (d) Long-Term Improvement: Over time, assessments can track trends and changes in employee sentiments regarding sense of belonging and reducing burnout.

Distributed remote settings, which is how OSS operates, present unique challenges in observing signs of burnout compared to traditional in-person work environments. While we can leverage data from software repositories to investigate potential burnout signals, the effectiveness of these techniques is not universal due to contextual factors not possible to collect via archival data. Cultural differences, language barriers, personality traits, communication styles, and other individual differences can significantly influence the detection and interpretation of burnout signals.

Identifying burnout signals is an open challenge for further investigation. Techniques such as mining software repositories, sentiment analysis, and Natural Language Processing can automate the detection of burnout signals by analyzing communication patterns within project repositories \cite{raman2020stress}. Comments on code commits, discussions on mailing lists, and interactions in project forums can provide insights into contributors' engagement levels and overall well-being.

However, the applicability and accuracy of these techniques may vary, requiring a deeper exploration of how different contexts impact the detection of burnout signs. Future research should focus on understanding how diverse cultural backgrounds, communication preferences, and other individual factors influence the manifestation of burnout in remote settings. This ongoing investigation will help refine these techniques to be more inclusive and effective, ultimately enabling project and community managers to better support contributors at risk of burnout by providing tailored interventions such as encouragement, workload redistribution, skill development resources, and facilitated breaks.

\vspace{-2mm}

\section{ACKNOWLEDGMENT}

This work is partially supported by the National Science Foundation under Grant Numbers 1900903 and 2303612.

\vspace{-2mm}

\bibliographystyle{IEEEtran}
\bibliography{references}

\begin{IEEEbiography}{Bianca Trinkenreich} is an Assistant Professor at Colorado State University. She researches social aspects of Software Engineering and CSCW. Contact her at bianca.trinkenreich@colostate.edu. Find more at biancatrink.github.io.
\end{IEEEbiography}

\begin{IEEEbiography}{Marco Gerosa} is a Full Professor at Northern Arizona University. He researches Software Engineering and CSCW. Contact him at marco.gerosa@nau.edu.
\end{IEEEbiography}

\begin{IEEEbiography}{Anita Sarma} is a Full Professor and Head of Research at Oregon State University. She researches inclusive technological support. Contact her at anita.sarma@oregonstate.edu.
\end{IEEEbiography}

\begin{IEEEbiography}{Igor Steinmacher} is an Assistant Professor at Northern Arizona University. His main topic of research is behavior in OSS communities. Contact him at igor.steinmacher@nau.edu.
\end{IEEEbiography}

\end{document}